# REVISITING THE JURASSIC GEOMAGNETIC REVERSAL RECORDED IN THE LESOTHO BASALT (SOUTHERN AFRICA)

Michel Prévot,[1] Neil Roberts,[2] John Thompson,[3] Liliane Faynot,[1] Mireille Perrin[1] and Pierre Camps[1]

[1] Laboratoire de Tectonophysique, CNRS-UM2, Université de Montpellier II, 34095 Montpellier Cedex 5, France. E-mail: Michel.Prevot@dstu.univ-montp2.fr
[2] Magnetic Resonance and Image Analysis Research Centre (MARIARC), University of Liverpool, P. O. Box 147, Liverpool, L69 3BX, UK
[3] Formerly at: Laboratoire de Géomagnétisme, CNRS and Université de Paris 6, 94107 Saint-Maur Cedex, France

Abbreviated title : JURASSIC GEOMAGNETIC REVERSAL

ABSTRACT
   We carried out a detailed and continuous paleomagnetic sampling of the reversed to normal geomagnetic transition recorded by some 60 consecutive flow units near the base of the Lesotho Basalt (183 ± 1 Ma). After alternating field or thermal cleaning the directions of remanence are generally well clustered within flow units. In contrast, the thermal instability of the samples did not allow to obtain reliable paleointensity determinations. The geomagnetic transition is incompletely recorded due to a gap in volcanic activity attested both by eolian deposits and a large angular distance between the field directions of the flows underlying or overlying these deposits. The transition path is noticeably different from that reported in the pioneer work of van Zijl et al. (1962). The most transitional Virtual Geomagnetic Poles are observed after the volcanic hiatus. Once continents are replaced in their relative position 180 Ma ago, the post-hiatus VGP cluster over Russia. However, two successive rebounds from that cluster are found, with VGP reaching repeatedly Eastern Asia coast. Thus, the VGP path is not narrowly constrained in paleolongitude. The decrease in intensity of magnetization as the field deviates from the normal or reversed direction suggests that the decrease in field magnitude during the reversal reached 80-90%. We conclude that although the reversal is of a dipole of much weaker moment than that which existed on average during Cenozoic time, the characteristics of the reversing geodynamo seem to be basically similar.

## 1. Introduction

   Brynjolfsson (1957) and Sigurgeirsson (1957) were probably the first workers to report the occurrence of lava flows with intermediate directions of remanence between two volcanic magnetozones of opposite polarity. They suggested that these intermediate directions, observed on Icelandic lava flows of Upper Cenozoic age, offer a record of the transitional state of the geomagnetic field when changing from one polarity state to the other. A few years later, from an extensive magnetic study of some 150 cores recovered along a 1,300 m thick section of the Early Jurassic Lesotho Basalt (southern Africa), van Zijl et al. (1962a and b) reported an impressive line of observations in favor of a geomagnetic origin of the progressive reversal of remanence in successive lava flows, as opposed to the self-reversal mechanism proposed by Néel (1951, 1952). In the Maseru area, these authors were able to describe a transition zone extending between 290 and 442m above the base of the lava sequence (Fig. 1) between underlying reversed flows and the overlying normal ones. The most important observations of van Zijl et al. (1962a) in favor of the genuine character of field reversal were as follows : (i) baked sediments exhibit the same remanence direction as the overlying flow, (ii) reversed lava flows cut by normally magnetized dikes become normally magnetized as one approaches the contact, and (iii) lava flows and sediments as well acquire a TRM parallel to the laboratory field direction whatever their NRM polarity. Moreover, in their second paper van Zijl et al. (1962b) were the first workers to try to determine the paleostrength of the geomagnetic field while reversing. Comparing the intensity of a total TRM acquired in the laboratory to that of NRM (both partially demagnetized by AF), they concluded that the field intensity decreased by a factor of about four to five during the reversal. This total TRM paleointensity method is no longer in use because it offers no possibility to detect the changes in TRM capacity occurring during laboratory heating at high temperature (650°C in van Zijl et al.'s (1962b) experiments).

The present study was undertaken with the aim of obtaining a more precise description of the Lesotho geomagnetic reversal. In the original study, not all the consecutive flows were sampled and only one (long) core was collected from each sampled flow. Apparently, this field reversal occurred during a period of long-term dipole low (Prévot et al., 1990; Perrin and Shcherbakov, 1997). According to Kosterov et al. (1997) the average field intensity during Early Jurassic was only one half of the Late Cenozoic value. While there are several well documented records of Upper Cenozoic reversals (Mankinen et al., 1985; Prévot et al., 1985 a and b; Chauvin et al., 1990; Kristjansson et Sigurgeirsson, 1993; Goguitchaichvili et al., 1999 ; Leonhardt et al., 2002), the Lesotho reversal is to date the only detailed volcanic record of a geomagnetic transition which occurred during a long period of dipole low. This gives a particular interest to this reversal, since it is not yet known whether the characteristics of geomagnetic reversals depend upon the magnetic moment of the dipole field that existed during the preceding and following epochs of stable polarity.

## 2. Sampling and laboratory procedures
*Geology and sampling*

The Lesotho Basalt (Fig. 2) is part of the widespread volcanism that occurred in Southern Africa in Lower Jurassic times. Two sections were studied by van Zijl et al. (1962a and b) : Bushmen's Pass (near Maseru) and Sani Pass. Because the later section is more weathered, we focussed our efforts on the Bushmen's Pass section. In the course of a generalized sampling of the reverse zone, a few flows were also collected near Rhodes, some 150 km S of Bushmen's Pass. Four of these flows were found to be transitional and the relevant data will be presented here.

In the Maseru area, four sections (from bottom to top : Y, Z, X, and R) were sampled along or in the vicinity of the road going up from Nazareth to Bushmen's Pass (Fig. 3). These sections stratigraphically overlap with each other. Unambiguous correlations from one section to the other could be made between flows or groups of flows using stratigraphic evidences and directions of remanence. Altogether, we think that our sampling was continuous between 2010m (base of Y) and 2350m (top of R). The successive flows were numbered upwards (Y7 to Y21; Z1 to Z10; X1 to X25; R1 to R12). A total of 56 distinct flows (785 cores) was sampled (4 to 8 cores per flow). The four transitional flows from the Rhodes area were collected at two sites within the bed of Bell river (lat. 30° 45' 37" S and long. 28° 02' 47" E for RH4 and RH5; lat. 30° 45' 49" S and long. 28° 03' 22" E for RH7 and RH8, collected near Naude's Neck monument).

Along the Bushmen's Pass composite section, the average flow thickness is approximately 6 m. Individual values vary from 1 to 20m. Typically, the flow base is characterized by vertically elongated vesicles, which differ markedly from the more spherical or horizontally flattened vesicles from the upper parts of flows. Ropy flow tops were occasionally observed. A major feature of the Lesotho Basalt sequence is the occurrence of a thin sedimentary level within the geomagnetic transition zone (van Zijl et al., 1962a). Along the Busmen's Pass sequence, a horizon of sedimentary lens, each up to 1.5m thick and several meters long, was observed near elevation 2160m. We were able to trace this discontinuous sedimentary layer over 1km along the Western slope of the Thaba-Tseka mountain (Fig. 3). We will see below that this horizon corresponds to a distinctive break in the geomagnetic record. There is no evidence for a regional tilting of the lava pile (Hargraves et al., 1997).

Fitch and Miller (1971, 1984) reported the first K-Ar and $^{39}Ar/^{40}Ar$ of the Lesotho Basalt at Buhmen's Pass. They concluded that this lava section erupted very rapidly at some time close to 193 Ma. More recent $^{39}Ar/^{40}Ar$ dating (Duncan et al., 1997), once recalculated using the standard hornblende monitor Mmhb-1 (523.4 Ma, astronomically tuned), provides a mean age of 183 ± 1 Ma for the Lesotho formation as a whole. Paleomagnetic data are compatible with an extremely rapid eruption of the total 1400m thick sequence. Considering that the reversal is recorded over a thickness of some 150m and that the mean duration for field reversal is of the order of 5,000-6,000 years (Kristjansson, 1985), the eruption of the Lesotho Basalt might have lasted for some 50,000 years or so.

Thus, the mean age of the lava pile is also the best radiometric estimate of the age of the reversal, which has therefore to be considered as Toarcien. According to the geomagnetic reversal time scale of Gradstein et al. (1994), the Lesotho reversal might be either 182.4 or 183.2 Ma old.

*Laboratory procedures*

We first determined the two-week viscosity index (Thellier and Thellier, 1944; Prévot, 1981) of one specimen from each core, with the exception of cores from X section, which were first magnetically analyzed at the Liverpool laboratory. Then another specimen from each core was progressively demagnetized by alternating fields. Several specimens from each flow were also demagnetized by stepwise heating in zero field in air. For each specimen, the direction of the characteristic remanent magnetization (ChRM) was calculated using principal component analysis (Kirschvink, 1980). A few flows were subjected to paleointensity experiments using the Thellier original method (Thellier and Thellier, 1959) and a high vacuum furnace. Unfortunately, due to the poor magnetic stability of most of the Lesotho lava flows during heating (Kosterov and Prévot, 1998), no really reliable paleointensity data could be obtained. However, NRM(T) directions calculated from these experiments were used as a substitute to data from standard thermal cleaning to provide the ChRM direction of the specimens studied for paleointensity purpose. This substitution is justified by the absence of noticeable CRM acquisition during paleointensity experiments (Kosterov and Prévot, 1998).

## 2. Paleomagnetic characteristics of samples

As previously observed (Prévot, 1981), the magnetic viscosity index v does not show a normal distribution (Fig. 4a) but rather a lognormal one (Fig. 4b). The mode on this figure corresponds to v = 5%. Considering the age of the rocks, this is a rather weak value, in fact close to the average viscosity index found for subaerial lava flows of Upper Cenozoic age (Prévot, 1981).

Examples of the orthogonal plots obtained from either AF or thermal progressive demagnetization are given in Fig. 5. As a general rule, a quite simple magnetic behavior is observed in the lowermost (reversed) and the uppermost (normal) lava flows (Fig. 5a). With the exception of a small VRM, a single component is observed. Thermal and AF demagnetization yield the same direction of ChRM although some deflection of the direction can be sometimes seen on the AF diagrams when approaching a completely demagnetized state.

For the median part of the composite section, the behavior of the samples is not always as straightforward. As shown by Fig. 5b, thermal cleaning appears more suitable than AF cleaning, even when the secondary component is small. It can be inferred from Fig. 5c that most of the difficulties encountered during AF cleaning is due to the acquisition of some parasitic magnetization (ARM or RRM) which progressively deflects the direction of remanence. In such specimens, thermal cleaning is necessary to reach an approximately zero-magnetization state in the specimen (Fig. 5c, specimen 86P133A).

Typically, the natural secondary components remain small in the flows from the transition zone and thermal treatment allows a precise determination of ChRM direction (Fig. 5d, specimen 86P482B). However, a few samples carry a quite large low temperature secondary component, possibly of viscous origin for the example shown in Fig. 5d (specimen 86P128A). In a restricted area around Bushmen's Pass, corresponding to the uppermost flows of the X section (above X22), several cores were found to carry a large IRM (in spite of the precautions taken in the field to avoid sites struck by lighting), which made it impossible to determine ChRM directions.

The flow average direction was in general calculated from the combination of the remanence directions of both thermally and AF cleaned specimens. Although the cleaning range used for principal component analysis was variable from specimen to specimen, the ranges 400-550°C or 20-50 mT are rather typical. The flow directions so obtained are listed in Table 1. The quality of the data is very good both in the reversed and normal zones (almost 95% of sampling sites with dispersion parameter k>100) but only fairly good in the transition zone (70% of sampling sites with dispersion parameter k>100).

## 3. Description of the composite geomagnetic transition record

Stratigraphic observations and paleomagnetic data both agree that the four sections Y, Z, X and Z slightly overlap, which allows reconstruction without any gap of the paleomagnetic behavior recorded by the Bushmen's Pass sequence. The composite record consists of 36 "distinct" paleomagnetic directions (Table 2 and Fig. 6) distinguished from each other using the method of Mankinen et al. (1985). Each of these directions defines a "paleomagnetic unit". One third of these paleomagnetic units are defined from several (up to six) consecutive lava flows yielding the same direction (overlapping $\alpha_{95}$ semi-angles) called "directional flow groups" by Mankinen et al. (1985). The limits of the transition zone were defined from the reversal angle (Prévot et al., 1985) of each paleomagnetic unit. The reversal angle of a paleomagnetic unit is the angular distance between the paleomagnetic direction of this unit and the direction of either the normal or reversed mean field, whichever is closer. The beginning (end) of the transition zone was chosen as corresponding to the first (last) flow from a sequence of at least two consecutive paleomagnetic units with reversal angle lying outside the $\theta_{95}$ semi-angle of individual directions. Following the results of the paleosecular variation study of the normal and reversed magnetozones of the Lesotho Basalt carried out by Kosterov and Perrin (1996), $\theta_{95}$ was taken equal to 24°.

The transition path is obviously discontinuous. A large angular gap is observed at the level of the sedimentary lens (elevation 2155m) between directions 11 and 12. This gap was first observed by Van Zijl (1962b) who also found it along the Sani Pass section, some 130 km East of Bushmen's Pass. The sedimentary lenses display cross stratification of eolian origin. Obviously, the paleomagnetic gap is due to a pause in volcanic activity. None of the transitional directions deviates by more than 54° from the steady field direction (Table 2), which suggests that the volcanic pause was coincident with the most transitional field configurations. The directional path presently obtained (Fig. 6) is more detailed and presents obvious differences from that reported by van Zijl et al's (1962b). These differences result probably from the fact that this pioneering study was based on a discontinuous sampling of the flow sequence, with generally only a single core from each sampled flow, and a moderate AF cleaning, sometimes insufficient to erase VRM or IRM due to lightning. Moreover, as we showed above, AF treatment of some samples can yield misleading results. The zigzag behavior of the transitional field directions described by these workers (Fig. 1) is not confirmed by our data. Instead, our record is compatible with rather gradual changes in direction during the two parts (pre and post gap) of the reversal. Furthermore, a rather complex pattern is observed during the final stages of the transition.

The VGP positions (Table 2) were first calculated as usual with Africa in its present position, then with Africa rotated back in its 180 Ma old position according to Morgan (1983). This paleoreconstruction assumes fixity of the Atlantic and Indian Ocean hotspot system as a whole. Figure 7 represents the paleoVGP path of the Lesotho reversal onto the 180 Ma old plate reconstruction proposed by Morgan (1983). The geographic paleopole used for that figure is the global paleomagnetic pole for the period 175-200 Ma as calculated by Prévot et al. (2000) from a selected dataset of paleomagnetic data from all continents obtained from magmatic rocks. Most of the transitional VGPs plot over Russia. However two successive "rebounds" (Watkins, 1969) are observed with extreme VGP positions close to the margin of the Eurasia and Pacific plates.

In the absence of direct paleointensity determinations, we tried to obtain a qualitative description of field paleostrength changes during this transition using cleaned remanence intensities as a proxy. This approach was already used by several authors (Dagley and Wilson, 1971; Kristjansson, 1985; Chauvin et al., 1990; Camps and Prévot, 1996). In the case of the Lesotho Basalt, the use of this indirect method seems reasonable considering the petrographic monotony of this volcanic suite (Cox, 1988) and the relative weakness of secondary magnetization's as compared to ChRM in most of the rock samples. Moreover, rather than using NRM intensity, we used the remanence intensity after cleaning either by a 10 mT alternating field or heating at 200°C. A decrease in remanence intensity is observed in the transition zone (Fig. 8), in qualitative agreement with the paleointensity results of van Zijl et al. (1962b) who suggested a four to five-fold diminution in field intensity. A more representative estimate of the specific magnetization decrease can be obtained by calculating the average magnetization intensity versus consecutive reversal angle intervals (Fig. 9). A rapid decrease is observed as soon as the reversal angle starts increasing. For reversal angles exceeding 20° the remanence intensity is reduced to approximately 15% of the value found for the directions laying within a 10° radius cone from the average steady field direction.

## 4. Discussion and conclusions

The present data confirm the suggestion of van Zijl et al (1962b) that, due to a pause in volcanic activity that occurred apparently just during the most transitional stages, the Lesotho reversal is incompletely recorded. No VGP very close to paleoequator is observed. Considering that the global paleopole is located at latitude 69.1 and longitude 311.4 for the period 175-200 Ma (Prévot et al., 2000), the lowest paleolatitude of transitional VGP is 20° (unit 21, Fig. 7). The fluctuations of the transitional field directions documented by the present study are however quite different from those reported by van Zijl et al (1962b). Also, they seem to have been rather gradual, with no compelling evidence in favor of the "jerky" behavior of the geomagnetic field advocated by those workers.

Both before and during the Lesotho reversal, several "directional flow groups" - each constituted of consecutive flows exhibiting the same direction (within experimental uncertainties) - are found. Such directional flow groups are of rather common occurrence in volcanic sequences (e.g. Mankinen et al., 1985; Hoffman, 1991; Mc Elhinny et al., 1996; Széréméta et al., 1999). In the case of the Lesotho sequence, such groups comprise up to 6 consecutive flows (direction 2) but, more commonly, only 3 to 5 (directions 5, 11, 22, 29). Directional groups can be interpreted either as multiple records of a single direction of a constantly changing geomagnetic field that are due to a brief outburst of volcanic activity (Mankinen et al., 1985) or as reflecting a standstill of the geomagnetic field direction (Hoffman, 1991). The first interpretation is supported by many observations on present volcanoes. However, the second interpretation can be preferred when significant changes in field paleostrenth are observed within a single directional group (Prévot et al., 1985b) or, alternatively, when stratigraphically distinct directional flow groups recurrently record a similar field direction. Such a case can be observed in the detailed record of the reversal found at Steens Mountain (Mankinen et al. 1985; Camps et al., 1999) in which directional flow groups 21 (3 consecutive flows) and 31 (7 consecutive flows) do yield the same transitional direction. Recently, a similar field recurrence was reported from another volcanic record of a mid-Miocene geomagnetic reversal found in Gran Canaria (Leonhart et al., 2002). In contrast, the transitional directions carried by the flow groups from the Lesotho record are all different from each other. Given the absence of recurrent paleomagnetic direction, we consider that these groups of flows are probably due to intense outpouring of several lava flows in a short interval of time rather than to successive standstills of the geomagnetic field.

Two successive N-I-N rebounds of the geomagnetic field towards intermediate VGPs are observed during the last stage of the Lesotho reversal after a first reestablishing of normal polarity (Fig. 7). Rebounds are commonly observed in detailed Cenozoic records of reversals (Mankinen et al, 1985; Chauvin et al., 1990; Kristjansson et Sigurgeirsson, 1993; Leonhardt et al., 2002). Paleomagnetic records of Cenozoic reversals obtained from sedimentary rocks indicate that transitional poles tend to fall along two longitude bands centered on America or Asia/Pacific boundary (Tric et al., 1991 ; Laj et al., 1991). The significance of this pattern is unclear. If not weighted, transitional poles from volcanic rocks from the past 16 Ma show no evidence for any longitudinal organization (Prévot and Camps, 1993). However, using debatable methods of normalization and weighting, Love (1998) observed a maximum in VGP longitude distribution between 60 and 90°E (Eastern Asia). Intriguingly, the rebound VGPs of the Lesotho reversal fall along the boundary between the Eurasia and Pacific plates. However, in the absence of other data from geomagnetic reversals with appropriate dates, we do not know if this observation is representative for the Jurassic.

The regularity of the decrease in averaged specific magnetization as the transitional field deviates from the normal/reverse field direction (Fig. 9) strongly suggests that this diminution reflects a progressive decrease of the geomagnetic paleostrength. Quite similar trends were reported from Cenozoic volcanic sequences in Iceland (Kristjansson, 1985) and Polynesia (Chauvin et al., 1990). Such trends are well fitted by a statistical model of the geomagnetic field in which the fluctuations of the non-axial dipole components are isotropic and independent from those of the axial dipole (Camps and Prévot, 1996). The trend in magnetization magnitude versus reversal angle (Fig. 9) suggests that the maximum decrease in the paleofield intensity might have reached 80-90% during the Lesotho reversal, which is comparable with the result obtained from direct paleointensity measurements for the well documented Cenozoic Steens Mountain reversal (Prévot et al., 1985 b). Taking into account the weakness of the local steady paleofied in Early Jurassic (24 ± 11μT according to Kosterov and Perrin, 1997), the predicted transitional field intensity would have been particularly low. The variation of the averaged specific magnetization as the field reverses (Fig. 8) suggests some asymmetry in the change in field magnitude from the pre to the post-transitional stages. A similar behavior is well substantiated by direct paleostrength determinations on several Cenozoic geomagnetic reversals (Prévot et al., 1985 b; Bogue and Paul, 1993; Quideller and Valet, 1996; Riisager and Abrahamsen, 2000).

Thus, although the paleomagnetic data for the Lesotho reversal are not as complete and accurate as for more recent reversals, they support the contention that the characteristics of geomagnetic reversals were similar in Late Cenozoic and Early Jurassic times. While the long-term averaged dipole moment of the Earth appears to have been twice weaker during Early Jurassic than during Late Cenozoic (Prévot et al., 1990; Perrin and Shcherbakov, 1997; Kosterov et al., 1997), the reversal process seems to have been basically unchanged.

## 5. Acknowledgements

We are indebted to the Institute of Southern African Studies, Lesotho and its Director, Prof. K. K. Prah, for welcoming three of us (M.P., N. R. and J. T. ) as Visiting Research Associates at the National University of Lesotho in 1986 and to J. S. van Zijl who then kindly joined us in the field along the Bushmen's Pass section. The paper benefited from reviewer comments from C. Laj, A. Mazaud, and an anonymous reviewer. The work was supported by CNRS-INSU (programs "ATP Noyau 1986" and "Terre Profonde 1991") and a personal fellowship to N. R. from the N.E.R.C.

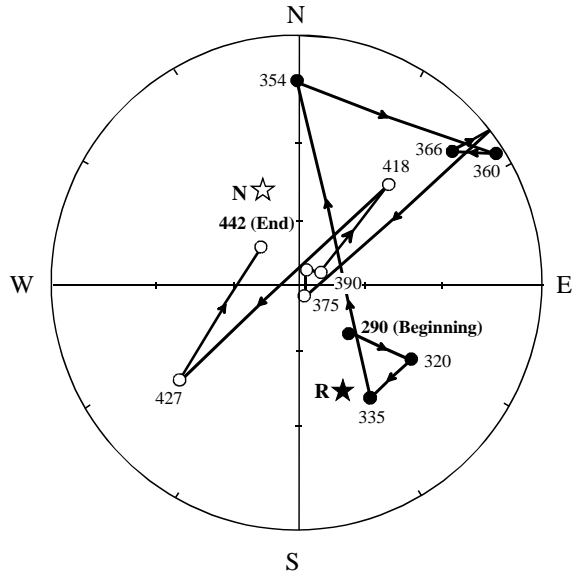

Fig. 1. Stereographic projection of the cleaned direction of remanence of successive flows from the Lesotho transition zone in Maseru area according to van Zijl et al. (1962a) (redrawn). The figures by the dots indicate the approximate flow elevation in meters above base of lava sequence.

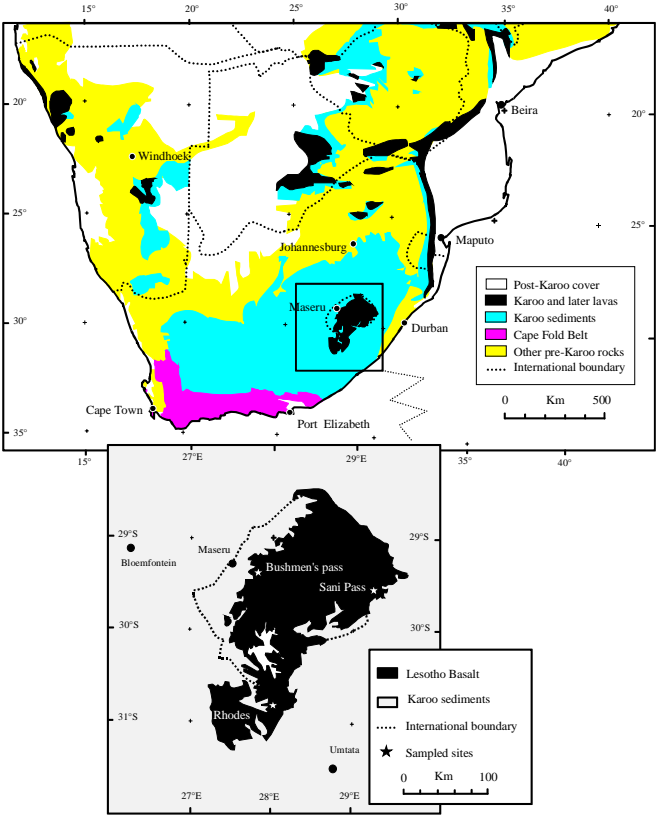

Fig. 2. The three main sampling localities of the Lesotho reversal. Maps redrawn from Kosterov and Perrin (1996).

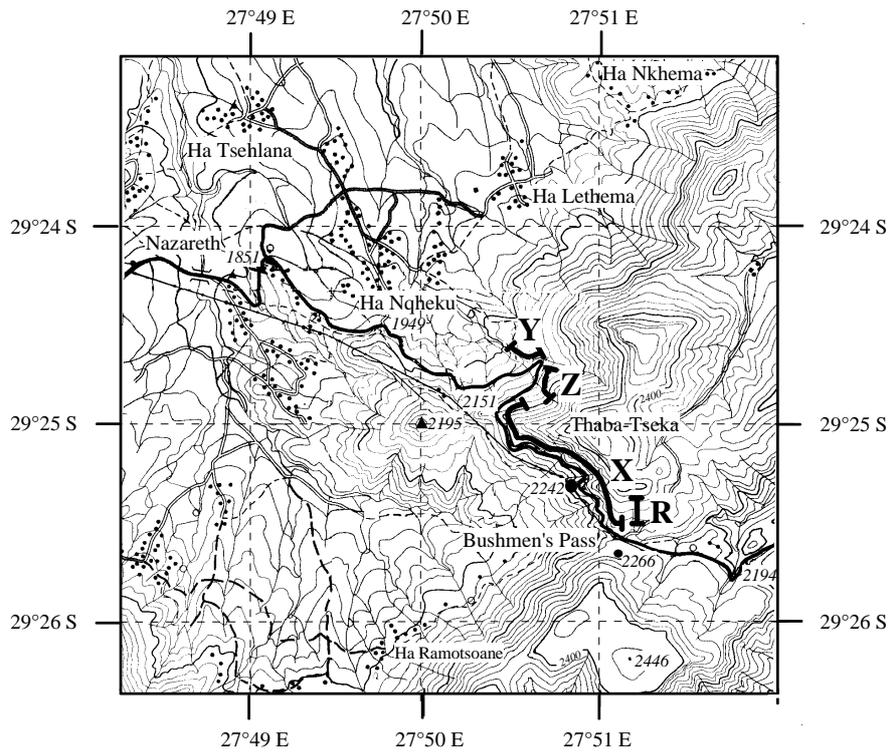

Fig. 3. Precise location of the four sections sampled by us between Nazareth and Bushmen's Pass (Maseru area).

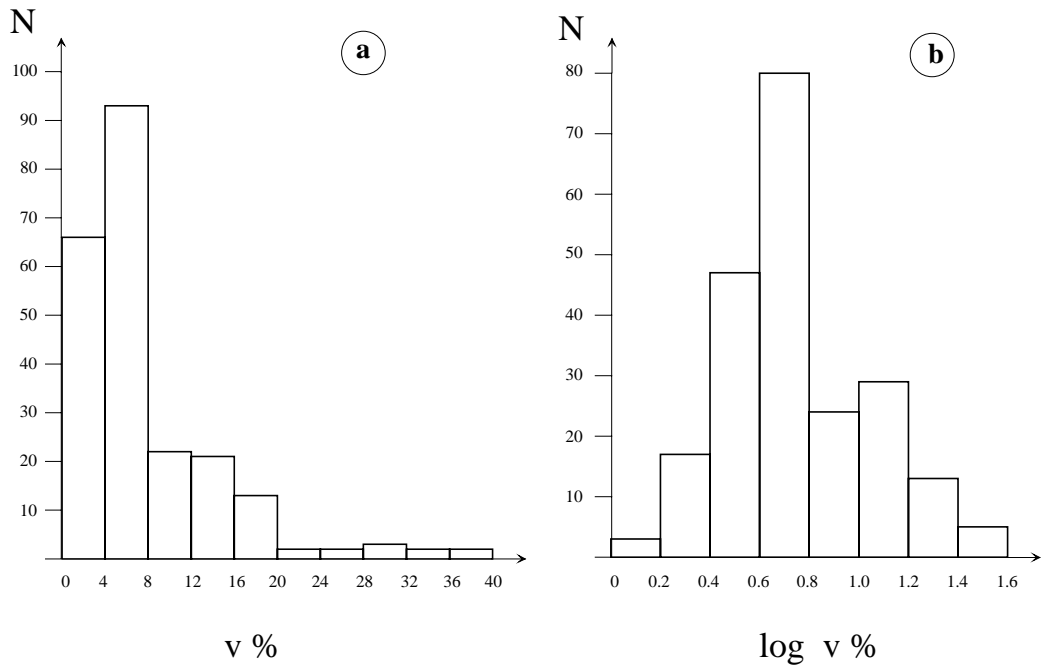

Fig. 4. Distribution of magnetic viscosity index (see text) from a total of 224 specimens.

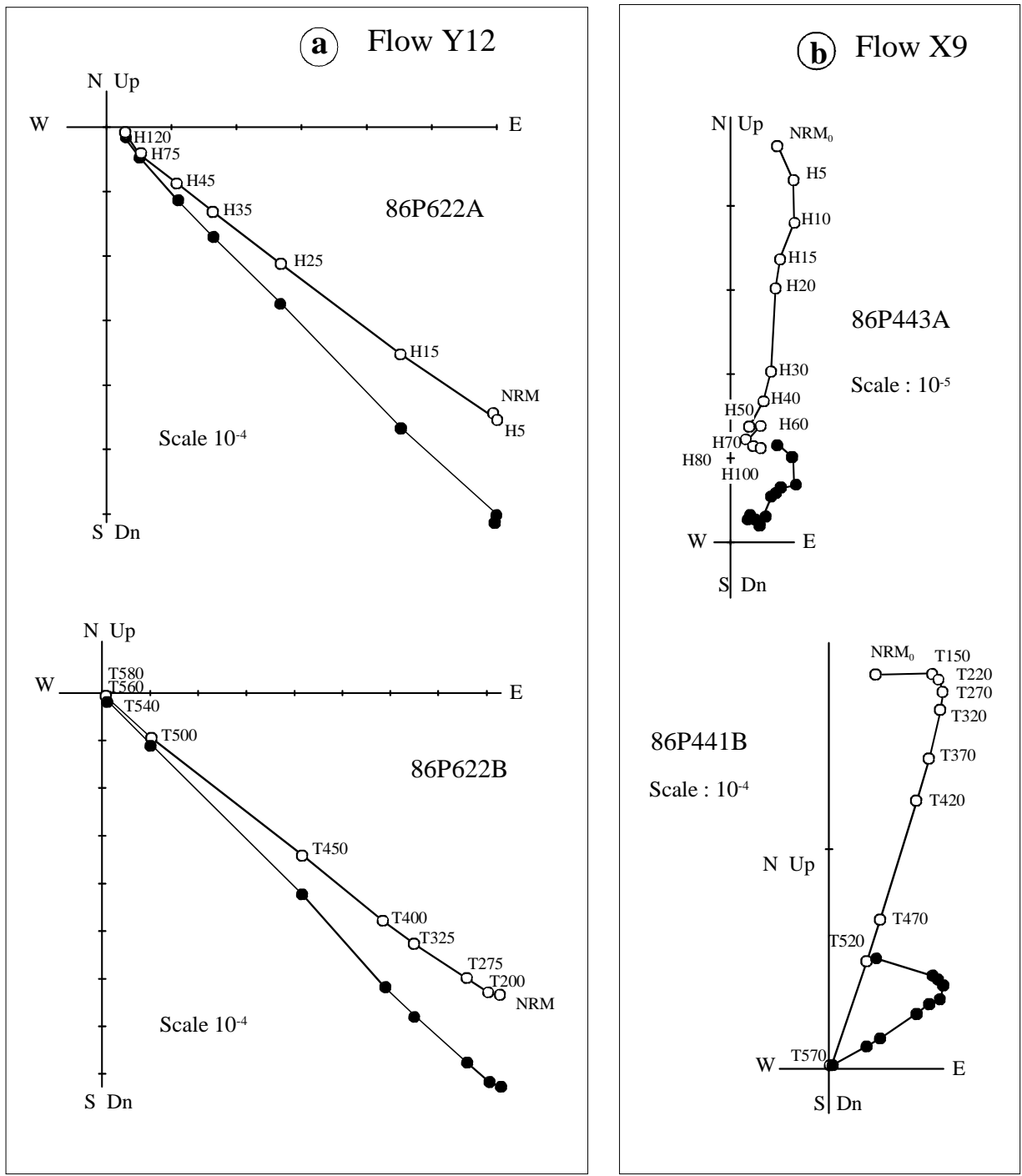

Fig. 5 (a to d). Examples of orthogonal thermal or alternating field demagnetization diagrams (stratigraphic coordinates) for a total of 8 specimens from one flow from the reversed zone (Y12) and three flows from the transition zone (lower part, Z10; middle part, X9; upper part, X15). Full (empty) circles represent data projected onto the (one) horizontal (vertical) plane. Scale bars correspond to specific magnetization (Am$^2$ kg$^{-1}$). Cleaning steps are labeled either T (temperature in °C) or H (peak field in mT).

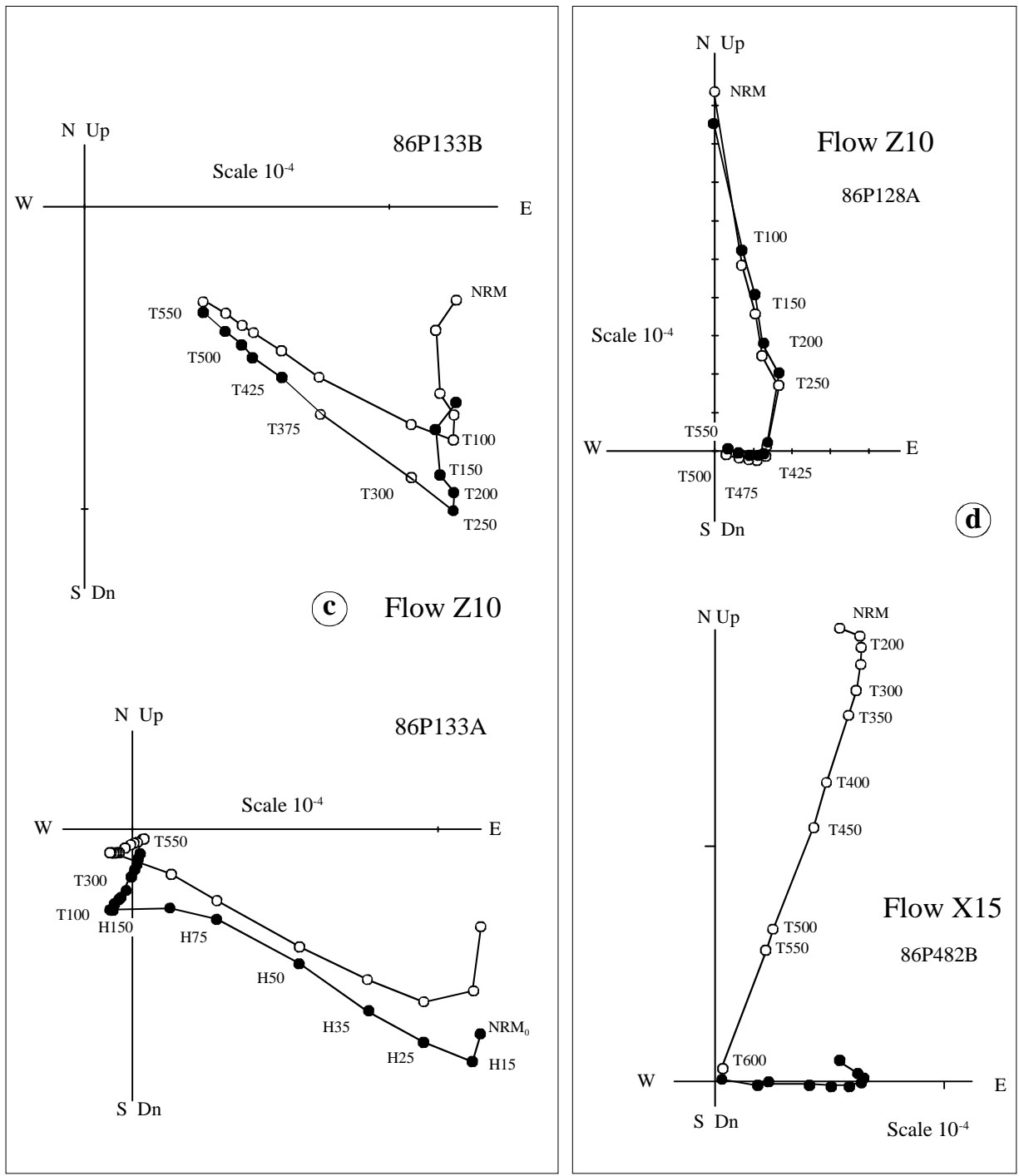

Fig. 5 (a to d). Examples of orthogonal thermal or alternating field demagnetization diagrams (stratigraphic coordinates) for a total of 8 specimens from one flow from the reversed zone (Y12) and three flows from the transition zone (lower part, Z10; middle part, X9; upper part, X15). Full (empty) circles represent data projected onto the (one) horizontal (vertical) plane. Scale bars correspond to specific magnetization (Am$^2$ kg$^{-1}$). Cleaning steps are labeled either T (temperature in °C) or H (peak field in mT).

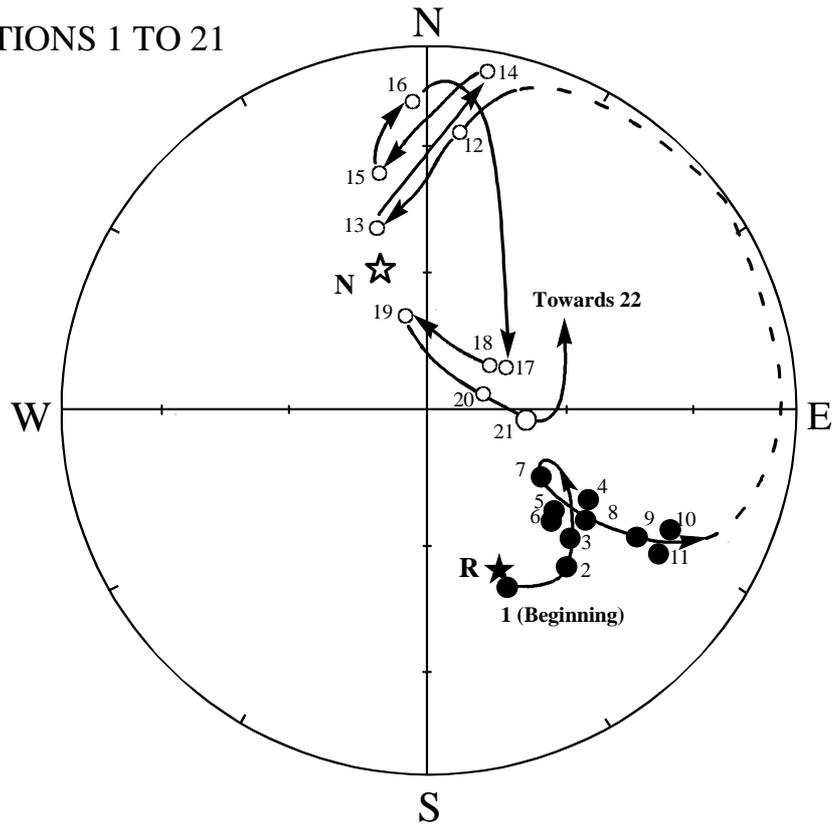

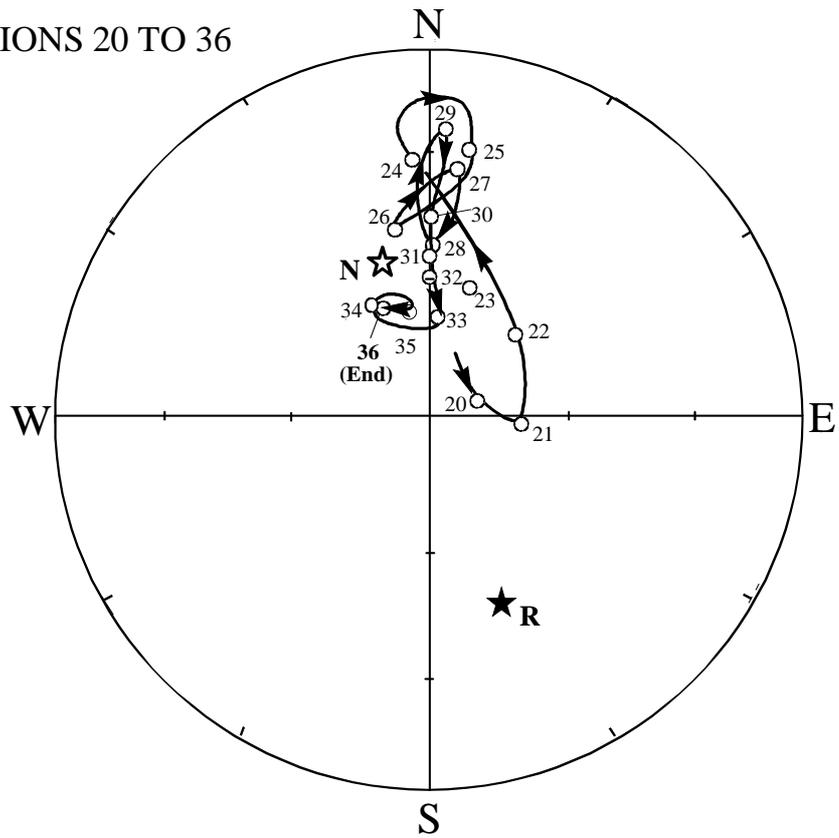

Fig. 6. Equal area projection of average ChRM directions of consecutive paleomagnetic units (flow or group of consecutive flows) numbered from the reversed unit 1 (base of composite section) to the normal unit 36 (top of composite section), as listed in Table 2.

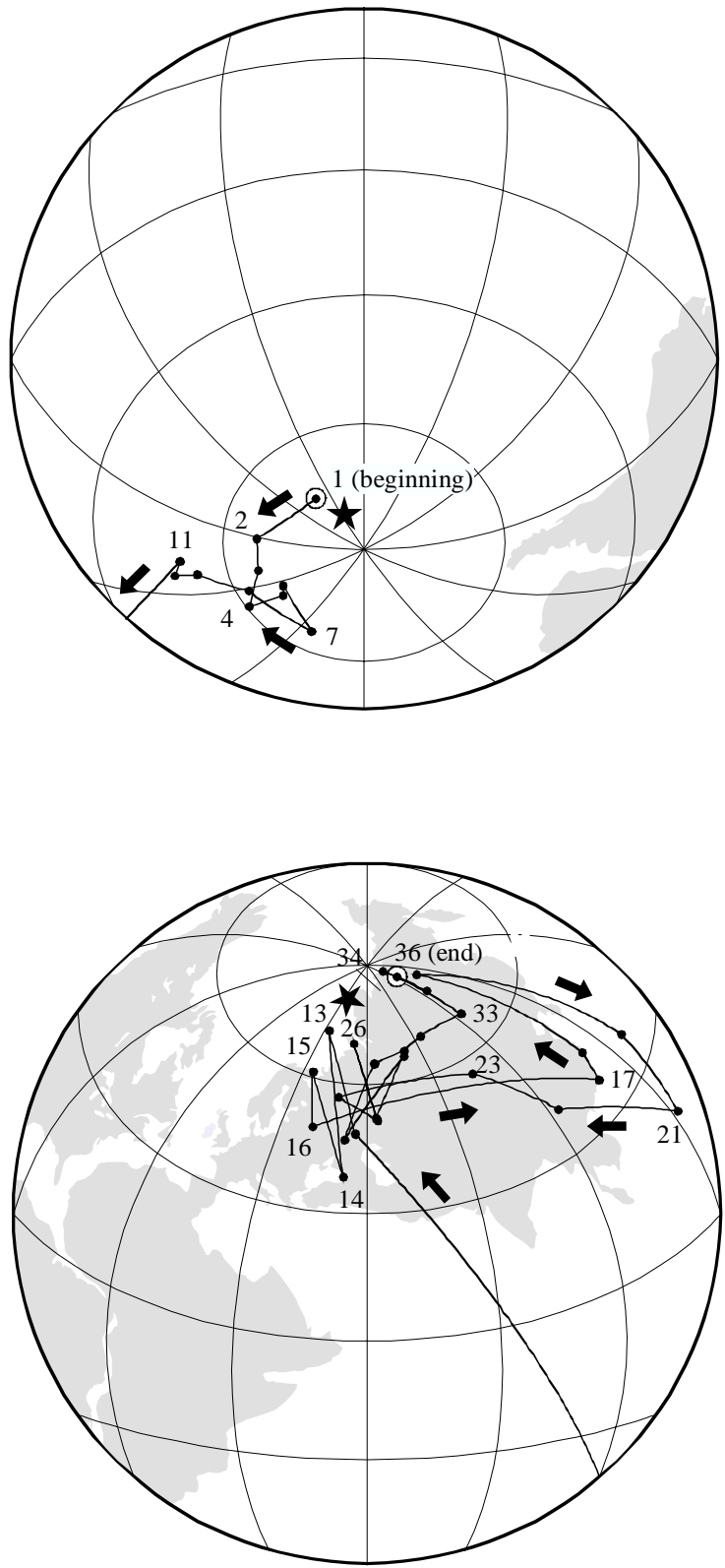

Fig. 7. Equal area projection of virtual paleopole positions during the Lesotho reversal as recorded along the Bushmen's Pass composite section. Main continents (with present contours) and VGPs have been rotated according to the 180 Ma old plate reconstruction proposed by Morgan (1983). The paleogeographic reference frame is pinned to the global paleopole position for the period 175-200Ma (Prévot et al., 2000). Star show the rotated position of the local paleomagnetic pole obtained by Kosterov and Perrin (1996) from normal and reversed sequences from the Lesotho Basalt. Projection poles are (30°N, 60°E) for the Northern hemisphere and (45°S, 240°E) for the Southern hemisphere.

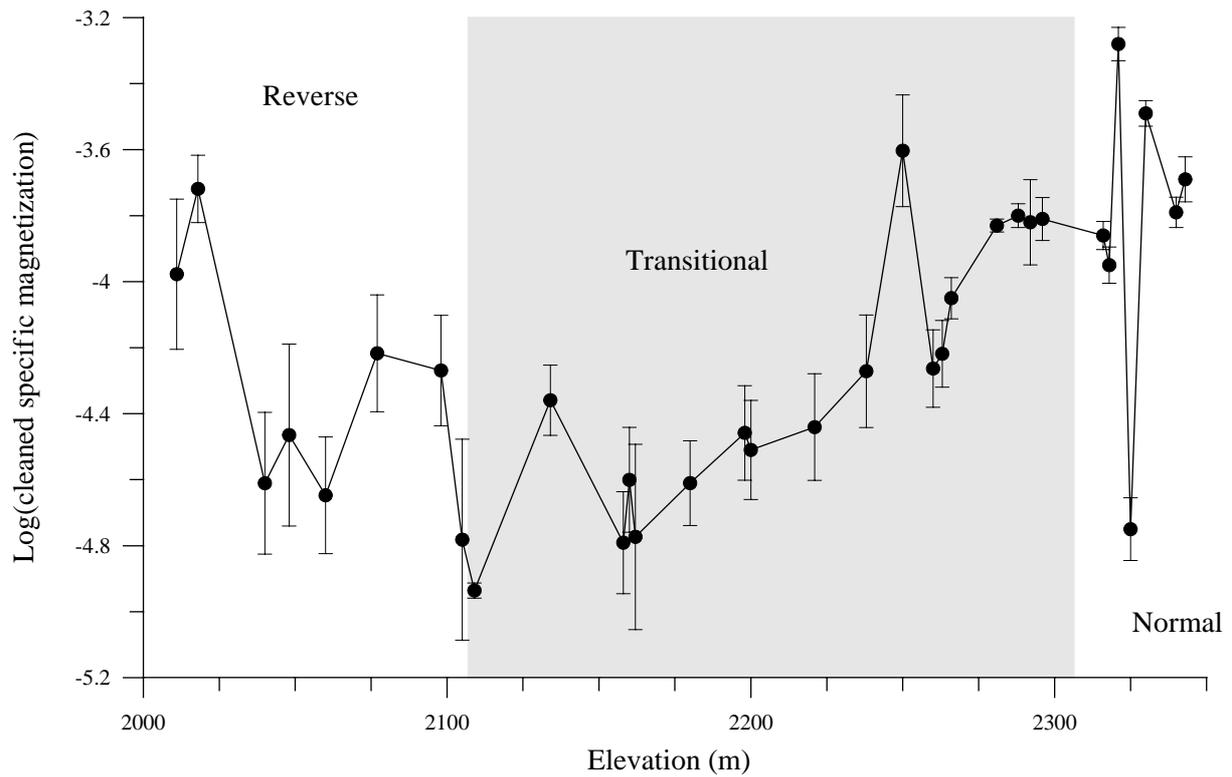

Fig. 8. Cleaned specific magnetization (logarithmic scale) of paleomagnetic units 1 to 36 (except 10) plotted in function of the elevation of the base of the unit.

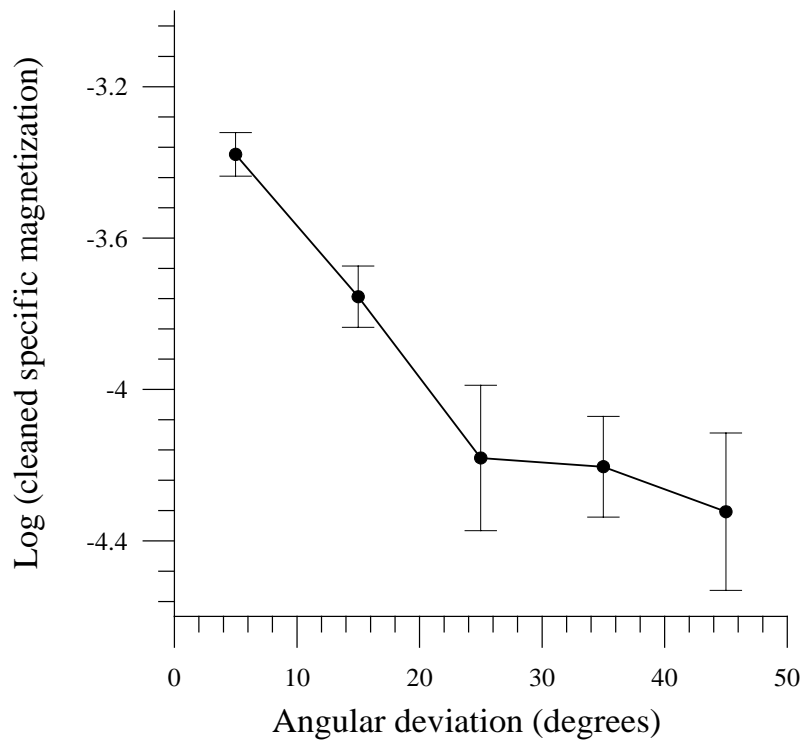

Fig. 9. Cleaned specific magnetization (logarithmic scale) of paleomagnetic units 1 to 36 (except 10) averaged over five adjacent 10° wide intervals of the reversal angle. By convention, the average magnetization is plotted at the middle of the corresponding interval of the reversal angle : 5°, 15°, 25°, 35° and 45°.

Table 1. Average ChRM directions and VGP coordinates of successive flows from Bushmen's Pass (sections Y, Z, X and R) and Rhodes sections. Paleomagnetic units (P. U.) are numbered from 1 to 36 (see text), elevation corresponds to flow base, N is the number of specimen directions used, I is inclination (degrees), D is declination (degrees), k is Fisher precision parameter, $\alpha_{95}$ is semi-angle of the 95 % confidence cone, PLA and PLO are the VGP latitude and longitude in degrees.

| Flow label | P. U. | Elevation (m) | N | I | D | k | $\alpha_{95}$ | PLA | PLO |
|---|---|---|---|---|---|---|---|---|---|
| **Bushmen's Pass (appr. 29.4S. 27.8E)** | | | | | | | | | |
| Y7 | 1 | 2011 | 5 | 47.2 | 157.7 | 60 | 10 | -70.5 | 115.4 |
| Y8 | 2 | 2017 | 6 | 40.9 | 141.6 | 231 | 4.4 | -55.2 | 118.5 |
| Y9 | 2 | 2020 | 10 | 42.3 | 138.5 | 153 | 3.9 | -52.9 | 115.3 |
| Y10B | 2 | 2021 | 7 | 42.1 | 136.0 | 389 | 3.1 | -50.6 | 114.5 |
| Y10A | 2 | 2022 | 4 | 45.5 | 142.0 | 1033 | 2.9 | -56.6 | 112.6 |
| Y11 | 2 | 2023 | 5 | 42.7 | 138.4 | 598 | 3.1 | -52.9 | 114.8 |
| Y13 | 2 | 2033 | 8 | 41.3 | 140.0 | 372 | 2.9 | -53.9 | 117.2 |
| Y14 | 3 | 2040 | 9 | 47.5 | 131.9 | 118 | 4.8 | -48.2 | 106.6 |
| Y15 | 4 | 2046 | 5 | 50.6 | 121.5 | 899 | 2.6 | -40.1 | 100.0 |
| Y16 | 4 | 2052 | 9 | 49.4 | 120.3 | 202 | 3.6 | -38.8 | 101.0 |
| Y17 | 5 | 2060 | 6 | 55.2 | 129.0 | 2106 | 1.5 | -47.0 | 95.6 |
| Y18 | 5 | 2065 | 3 | 52.0 | 131.2 | 178 | 9.3 | -48.4 | 100.5 |
| Y19 | 5 | 2072 | 8 | 55.8 | 127.1 | 317 | 3.1 | -45.6 | 94.4 |
| Y20 | 6 | 2076 | 7 | 53.9 | 130.7 | 140 | 5.1 | -48.3 | 97.7 |
| Y21 | 6 | 2081 | 5 | 53.1 | 133.9 | 661 | 3.0 | -50.8 | 99.4 |
| Z1 | 6 | 2095 | 6 | 54.7 | 130.4 | 538 | 2.9 | -48.1 | 96.5 |
| Z2 | 7 | 2099 | 7 | 61.9 | 117.6 | 543 | 2.6 | -39.1 | 84.2 |
| Z3 | 7 | 2101 | 8 | 62.5 | 122.2 | 763 | 2.0 | -42.4 | 83.6 |
| Z4 | 8 | 2105 | 7 | 48.0 | 125.0 | 63 | 7.6 | -42.5 | 103.9 |
| Z5 | 9 | 2109 | 6 | 34.4 | 121.3 | 225 | 4.5 | -36.0 | 114.9 |
| Z6 | 10 | 2112 | 4 | 28.8 | 117.3 | 621 | 3.7 | -31.0 | 116.9 |
| Z7 | 11 | 2132 | 7 | 28.9 | 120.8 | 254 | 3.8 | -34.1 | 118.5 |
| Z9 | 11 | 2143 | 4 | 27.7 | 120.4 | 848 | 3.2 | -33.4 | 119.1 |
| Z10 | 11 | 2145 | 11 | 27.8 | 121.1 | 335 | 2.5 | -34.1 | 119.4 |
| X1 | 11 | 2145 | 3 | 28.6 | 122.8 | 338 | 6.7 | -35.8 | 119.7 |
| X2 | 11 | 2150 | 3 | 24.0 | 121.6 | 179 | 9.2 | -33.5 | 122.0 |
| X3 | 11 | 2154 | 5 | 28.2 | 122.8 | 353 | 4.1 | -35.7 | 120.0 |
| Sedimentary lens | | | | | | | | | |
| X4 | 12 | 2157 | 5 | -23.8 | 5.5 | 75 | 8.9 | 72.3 | 45.7 |
| X5 | 13 | 2160 | 5 | -46.1 | 343.3 | 53 | 10.6 | 75.2 | 301.3 |
| X6 | 14 | 2162 | 8 | -5.6 | 9.3 | 81 | 6.2 | 62.0 | 47.9 |
| X7 | 15 | 2180 | 3 | -33.0 | 347.8 | 448 | 5.8 | 74.1 | 340.8 |
| X8 | 16 | 2197 | 6 | -15.6 | 356.5 | 275 | 4.0 | 68.3 | 18.4 |
| X9 | 17 | 2200 | 5 | -70.9 | 58.6 | 184 | 5.7 | 41.5 | 167.4 |
| X10 | 18 | 2212 | 4 | -73.6 | 50.0 | 376 | 4.7 | 45.0 | 174.5 |
| X11 | 19 | 2218 | 3 | -67.6 | 344.8 | 318 | 6.9 | 66.0 | 232.0 |
| X12 | 20 | 2221 | 6 | -77.1 | 64.5 | 101 | 6.7 | 37.0 | 179.7 |
| X13 | 20 | 2226 | 7 | -78.8 | 77.3 | 115 | 5.7 | 31.8 | 182.8 |
| X14 | 21 | 2238 | 5 | -71.5 | 91.9 | 73 | 9.0 | 23.1 | 170.6 |
| X15 | 21 | 2245 | 6 | -68.0 | 94.6 | 243 | 4.3 | 19.7 | 166.1 |
| X16 | 22 | 2250 | 12 | -63.7 | 43.1 | 171 | 3.3 | 52.8 | 155.2 |
| X17 | 22 | 2254 | 8 | -63.5 | 46.3 | 454 | 2.6 | 50.6 | 154.3 |
| X18 | 22 | 2256 | 8 | -61.4 | 51.4 | 284 | 3.3 | 47.1 | 150.0 |
| X19 | 23 | 2260 | 8 | -59.5 | 19.2 | 54 | 7.6 | 70.9 | 157.8 |
| X20 | 24 | 2263 | 9 | -31.2 | 356.7 | 76 | 5.9 | 77.1 | 13.5 |
| X21 | 25 | 2266 | 6 | -33.1 | 7.5 | 49 | 9.7 | 76.7 | 60.6 |
| X23A | 26 | 2281 | 5 | -47.7 | 350.2 | 64 | 9.6 | 81.4 | 299.5 |
| X23B | 27 | 2287 | 3 | -38.3 | 7.1 | 71 | 14.8 | 79.9 | 68.6 |
| X24 | 27 | 2288 | 5 | -30.2 | 7.0 | 30 | 14.0 | 75.3 | 55.3 |
| X25 | 28 | 2292 | 5 | -53.1 | 1.9 | 117 | 7.1 | 85.4 | 187.5 |
| X26 | 29 | 2316 | 3 | -14.4 | 4.7 | 16 | 31.8 | 67.5 | 40.0 |
| | | | | | | | | | |
| R1 | 29 | 2296 | 3 | -18.9 | 6.5 | 277 | 7.4 | 69.4 | 46.3 |
| R2 | 29 | 2301 | 4 | - 02 0 | 5.3 | 146 | 7.6 | 70.2 | 43.3 |
| R3 | 29 | 2306 | 4 | -22.8 | 4.5 | 151 | 7.5 | 72.0 | 42.2 |
| R4 | 29 | 2311 | 4 | -19.8 | 0.6 | 212 | 6.3 | 70.8 | 29.6 |
| R5 | 30 | 2316 | 5 | -46.8 | 357.7 | 209 | 5.3 | 87.6 | 331.4 |
| R6 | 31 | 2318 | 5 | -54.3 | 0.9 | 708 | 2.9 | 84.5 | 200.0 |
| R7 | 32 | 2321 | 5 | -59.2 | 1.1 | 1363 | 2.1 | 79.4 | 203.2 |
| R8 | 33 | 2325 | 5 | -67.6 | 4.8 | 228 | 5.1 | 68.6 | 199.4 |
| R9 | 34 | 2330 | 5 | -63.9 | 337.9 | 125 | 6.9 | 66.3 | 248.7 |
| R10 | 35 | 2340 | 5 | -66.7 | 351.6 | 1339 | 2.1 | 69.1 | 223.3 |
| R11 | 36 | 2343 | 3 | -65.9 | 335.0 | 428 | 6.0 | 63.2 | 246.4 |
| R12 | 36 | 2350 | 4 | -64.1 | 346.8 | 186 | 6.8 | 70.6 | 236.4 |
| **Rhodes (30.76°S. 28.05°E)** | | | | | | | | | |
| RH4 | 14 | 1965 | 5 | -5.6 | 4.3 | 225 | 5.1 | 61.7 | 37.2 |
| RH5 | 14 | 1970 | 10 | -8.8 | 6.3 | 220 | 3.3 | 63.0 | 42.0 |
| RH7 | around 22? | 1975 | 5 | -51.9 | 59.2 | 2673 | 1.5 | 40.3 | 136.5 |
| RH8 | around 22? | 1980 | 6 | -44.4 | 33.0 | 54 | 9.2 | 60.7 | 117.2 |

| Paleomag. Unit | Flow number(s) | N | I | D | k | α95 | δ | PLA | PLO | PPLA | PPLO |
|---|---|---|---|---|---|---|---|---|---|---|---|
| *Reversed magnetozone* | | | | | | | | | | | |
| 1 | Y7 | 5 | 47,2 | 157,7 | 60 | 9,9 | 6,2 | -70.5 | 115.4 | -59,3 | 160,9 |
| 2 | Y8, Y9, Y10A, Y10B, Y11, Y13 | 41 | 42,3 | 139,2 | 273 | 1,4 | 17,0 | -53.5 | 115.6 | -46,9 | 141,3 |
| 3 | Y14 | 9 | 47,5 | 131,9 | 118 | 4,8 | 17,8 | -48.2 | 106.6 | -45,8 | 130,2 |
| 4 | Y15, Y16 | 14 | 49,8 | 120,7 | 289 | 2,3 | 23,5 | -39.3 | 100.7 | -40,7 | 118,6 |
| 5 | Y17, Y18, Y19 | 17 | 54,9 | 12 8,5 | 375 | 1,8 | 17,5 | -46.6 | 95.9 | -48,7 | 120,1 |
| 6 | Y20, Y21, Z1 | 18 | 53,9 | 131,5 | 276 | 2,1 | 15,9 | -48.9 | 97.8 | -49,8 | 123,8 |
| 7 | Z2, Z3 | 15 | 62,2 | 120,0 | 607 | 1,6 | 21,9 | -40.8 | 84.0 | -48,7 | 104,4 |
| 8 | Z4 | 7 | 48,0 | 125,0 | 63 | 7,6 | 21,7 | -42.5 | 103.9 | -42,1 | 123,6 |
| *Transitional magnetozone* | | | | | | | | | | | |
| 9 | Z5 | 6 | 34,4 | 121,3 | 225 | 4,5 | 32,3 | -36.0 | 114.9 | -32,2 | 128,7 |
| 10 | Z6 | 4 | 28,8 | 117,3 | 621 | 3,7 | 38,8 | -31.0 | 116.9 | -27,0 | 127,8 |
| 11 | Z7, Z9, Z10, X1, X2, X3 | 35 | 27,7 | 121,5 | 340 | 1,3 | 37,4 | -34.4 | 119.7 | -29,0 | 131,9 |
| | Sedimentary lens | | | | | | | | | | |
| 12 | X4 | 5 | -23,8 | 5,5 | 75 | 8,9 | 35,9 | 72.3 | 45.7 | 49,4 | 31,7 |
| 13 | X5 | 5 | -46,1 | 343,3 | 53 | 10,6 | 7,9 | 75.2 | 301.3 | 60,7 | 350,5 |
| 14 | X6 | 8 | -5,6 | 9,3 | 81 | 6,2 | 54,0 | 62.0 | 47.9 | 39,5 | 36,0 |
| 15 | X7 | 3 | -33,0 | 347,8 | 448 | 5,8 | 21,5 | 74.1 | 340.8 | 53,2 | 3,5 |
| 16 | X8 | 6 | -15,6 | 356,5 | 275 | 4,0 | 40,4 | 68.3 | 18.4 | 44,7 | 18,8 |
| 17 | X9 | 5 | -70,9 | 58,6 | 184 | 5,7 | 37,6 | 41.5 | 167.4 | 58,0 | 144,2 |
| 18 | X10 | 4 | -73,6 | 50,0 | 376 | 4,7 | 34,6 | 45.0 | 174.5 | 63,5 | 150,4 |
| 19 | X11 | 3 | -67,6 | 344,8 | 318 | 6,9 | 14,5 | 66.0 | 232.0 | 78,8 | 303,0 |
| 20 | X12, X13 | 13 | -78,1 | 71,0 | 113 | 3,9 | 38,7 | 34.2 | 181.4 | 55,2 | 167,8 |
| 21 | X14, X15 | 11 | -69,6 | 93,5 | 123 | 4,1 | 48,4 | 21.2 | 168.1 | 39,4 | 156,7 |
| 22 | X16, X17, X18 | 28 | -63,0 | 46,5 | 226 | 1,8 | 35,2 | 50.5 | 153.3 | 60,3 | 118,5 |
| 23 | X19 | 8 | -59,5 | 19,2 | 54 | 7,6 | 22,9 | 70.9 | 157.8 | 73,0 | 75,6 |
| 24 | X20 | 9 | -31,2 | 356,7 | 76 | 5,9 | 25,8 | 77.1 | 13.5 | 53,6 | 17,8 |
| 25 | X21 | 6 | -33,1 | 7,5 | 49 | 9,7 | 33,9 | 76.7 | 60.6 | 54,8 | 35,5 |
| 26 | X23A | 5 | -47,7 | 350,2 | 64 | 9,6 | 9,4 | 81.4 | 299.5 | 64,1 | 1,8 |
| 27 | X23B, X24 | 8 | -33,3 | 7,0 | 39 | 9,0 | 28,5 | 77.1 | 59.0 | 55,0 | 34,6 |
| 28 | X25 | 5 | -53,1 | 1,9 | 117 | 7,1 | 13,9 | 85.4 | 187.5 | 70,7 | 25,7 |
| 29 | X26, RI, R2, R3, R4 | 18 | -19,2 | 4,3 | 70 | 3,9 | 35,6 | 70.1 | 40.3 | 46,9 | 29,7 |

*Normal magnetozone*

| | | | | | | | | | | |
|---|---|---|---|---|---|---|---|---|---|---|
| 30 | R5 | 5 | -46,8 | 357,7 | 209 | 5,3 | 16,8 | 87.6 | 331.4 | 64,8 | 17,3 |
| 31 | R6 | 5 | -54.3 | 0,9 | 708 | 2,9 | 13,2 | 84.5 | 200.0 | 71,8 | 23,0 |
| 32 | R7 | 5 | -59,2 | 1,1 | 1363 | 2,1 | 13,7 | 79.4 | 203.2 | 76,9 | 22,4 |
| 33 | R8 | 5 | -67,6 | 4,8 | 228 | 5,1 | 18,9 | 68.6 | 199.4 | 87,1 | 57,6 |
| 34 | R9 | 5 | -63,9 | 337,9 | 125 | 6,9 | 10,5 | 66.3 | 248.7 | 72,4 | 312,3 |
| 35 | RI0 | 5 | -66,7 | 351,6 | 1339 | 2,1 | 14,8 | 69.1 | 223.3 | 82,2 | 321,4 |
| 36 | RII,RI2 | 7 | -65,0 | 342,0 | 214 | 4,1 | 11,7 | 67.5 | 241.3 | 75,6 | 313,5 |

Table 2. Average ChRM directions and VGP coordinates (PLA, PLO) of consecutive paleomagnetic units numbered from 1 (lowermost unit) to 36 (uppermost unit). Symbols are similar to those of Table 1 with in addition the reversal angle δ (Prévot et al., 1985b) in degrees, and the VGP paleogeographic coordinates, paleolatitude (PPLA) and paleolongitude (PPLO), calculated from the position of the African plate 180 Ma ago according to Morgan (1983) and the global paleopole position calculated by Prévot et al. (2000) for the period 175-200 Ma.